\RequirePackage{fix-cm}

\documentclass[twocolumn,epjc3]{svjour3} 

\smartqed

\usepackage{listings}
\usepackage{enumitem}

\usepackage[utf8]{inputenc}
\usepackage{amsmath}
\usepackage{graphicx}

\usepackage[colorlinks=true,linkcolor=black,anchorcolor=black,citecolor=black,filecolor=black,menucolor=black,runcolor=black,urlcolor=black]{hyperref}

\usepackage{tikz}
\usetikzlibrary{automata,positioning}
\usetikzlibrary{shapes,arrows,chains}
\usetikzlibrary[calc]
\begin{document}

\tikzstyle{line} = [draw, -latex']

\title{A biased MC for muon production for beam-dump experiments}

\author{Stefan Ghinescu \and Babette D\"obrich \and Elisa Minucci \and Tommaso Spadaro}
\dedication{To be submitted to EPJ C}
\institute{S. Ghinescu \at
              Horia Hulubei National Institute of Physics and Nuclear Engineering, 30 Reactorului,
              P.O. Box MG 6, 077125-Magurele, Romania \\
              \email{stefan.alexandru.ghinescu@cern.ch} 
           \and
           B. D\"obrich and E. Minucci \at
              CERN, Esplanade des Particules 1, 1211 Geneva 23, Switzerland \\
              \email{babette.dobrich@cern.ch} \\
              \email{elisa.minucci@cern.ch} 
              \and
              T. Spadaro \at
              Laboratori Nazionali di Frascati INFN, Via E. Fermi 40, 00044 Frascati, Italy \\
              \email{tommaso.spadaro@cern.ch} 
              }
\date{}

\maketitle
\begin{abstract}
    The search for feebly-interacting new-physics particles in the MeV-GeV mass range often involves high-intensity beams dumped into thick heavy targets. The challenge of evaluating the expected backgrounds for these searches from first principles is limited by the CPU time needed to generate the shower induced by the primary beam. We present a  Monte Carlo biasing method allowing a three orders of magnitude increase in the efficiency for the simulation of the muon production in a 400~GeV$/c$ proton beam-dump setup. At the same time, this biasing method is maintaining nearly every feature of a simulation from first principles. 
\end{abstract}

\section{Introduction}

Realising more clearly the possibility that new particles beyond the Standard Model (SM) could be found at low mass and feeble couplings rather than at the highest achievable energies, has attracted the attention of many new experimental proposals. It has also lead to efforts to extend existing experiments in such a direction. The fact that many particle interactions are needed to observe possibly associated physics events puts such experiments under the umbrella of the ``intensity frontier''. In the `PBC frame-work' at CERN~\cite{Jaeckel:2020dxj}, initiatives in this context are, among else, the SHiP experiment~\cite{Anelli:2015pba}, the NA62~\cite{NA62:2017rwk} beam dump operation (NA62-BD), NA64~\cite{NA64} and FASER~\cite{FASER}. Other initiatives have started outside CERN, too, such as SeaQuest at FNAL~\cite{SeaQuest}.
While  differing vastly in the details of their implementation, these experiments have in common that an intense particle beam is shot to interact with a `heavy' target material. Rarely, such an interaction could create a low-mass (typically $<1$~GeV) exotic particle, very feebly coupled to SM particles, whose implications (decay or missing energy) are investigated after layers (typically many tens of meters) of `shielding'. This shielding is supposed to absorb most of the products from known physics processes, while letting the feebly interacting new-physics particles which are searched for, pass. The scarceness of the sought-after new-physics processes poses a challenge not only to the detection but also to the simulation of the experiment. Typically, on the order of $10^{18}$ or more primary particles are made to interact for proton-dump-experiments. Thus, in principle, Standard Model Processes that could constitute a background to a new-physics search have to be understood and simulated at that level, which seems an unfeasible challenge. 

The concrete problem we tackled is the simulation of the muon production after a high-energy proton beam is absorbed into a thick target. The generated muons can induce a variety of background to the searches for visible decays of feebly-interacting particles.
If a thick absorber made of a high-Z material is considered, the yield can be as low as $5\times 10^{-4}$ single muons above 10~GeV per proton, making a brute-force simulation quite inefficient CPU-wise. In the context of NA62-BD, a past approach to tackle this problem has been performed scoring the muons at the downstream face of the absorber and parametrizing the distributions obtained in terms of momentum, position, and direction~\cite{Rosenthal}. The parametrization was used as a particle gun for further simulation of the downstream beam line and experimental apparatus. In the context of the SHiP project, a more sophisticated approach using generative adversarial neural networks has been developed to define the particle gun~\cite{SHiP_GAN}. Both methods allow a reduction of CPU time budget per muon of five to six orders of magnitude. From the reliability point of view, both approaches critically depend on the statistics of the sample used to define or train the parametrization. The method here described constitute a decisive progress. Using a biasing simulation technique, it allows a dramatic boost of the statistical power with respect to a simulation from first principles with virtually no loss of information. The simulation from first principles, without any biasing applied, is called \textit{analogue} in this document.\\

The method described can be directly adapted to the simulation of high-intensity neutrino beams, if the appropriate parent particles are considered. The flux simulations at both the near and far detectors would benefit from the dramatic gain in CPU time efficiency, obtained while maintaining the principles of the hadronic shower model used. Moreover, the same biasing concepts can be applied to correctly evaluate the emission of exotic particles from every stage of the shower initiated by the beam particle.\\

\section{Generalities about biasing}

The efficiency of a Monte Carlo (MC) simulation can be defined as
\begin{equation}
    \label{eq:Biasing-FOM}
    \epsilon \propto \frac{1}{\sigma^2 T},
\end{equation}
where $\sigma$ and $T$ refer to the variance of the yield of interesting events (containing at least 1 muon) per event and the CPU time needed to simulate 1 full event, respectively. Note that this quantity is independent on the number of events simulated. The variance reduction (biasing) method described in this note has the remarkable feature of increasing the muon yield (thus lowering $\sigma$) by orders of magnitude, while only marginally increasing the CPU time $T$.
As with any biasing technique, special care is required in the implementation such that the physics simulated under biasing replicates as much as possible the analogue Monte Carlo (MC). Since the dominant muon sources are produced in all stages of the hadronic shower development, the biasing method must preserve the shape and composition of the analogue shower, i.e. there can be no modification to the number or kinematics of the particles created in an analogue event.

In the next section we give a heuristic description of the biasing method. Then, we illustrate  a concrete implementation of this method for a simple geometry, employing GEANT4~\cite{GEANT4} as simulation toolkit, since it allows for straightforward  usage of user-defined biasing schemes.

\section{Setup and algorithm description}
\subsection{GEANT4 configuration}
The geometry setup used to test the biasing algorithm is a simplification of the NA62 experiment absorber~\cite{NA62:2017rwk} (TAX) consisting of 2 blocks of copper followed by 6 blocks of iron. Each block has dimensions 0.78, 1.2, and 0.415~m in the $x$, $y$, and $z$ directions, respectively, with the $z-$axis oriented along the beam-line.  The primary particles are protons of $400~\mathrm{GeV/c}$ momentum moving along the $z$-axis, with a pencil beam profile, and the impact point is at $(0,0)$ in the $(x,y)$ plane. The overall TAX thickness corresponds to approximately 19 proton interaction lengths.

To the purpose of comparing biased with analogue muons, we have defined a scoring plane at the downstream face of the absorber.
We used the  \verb|FTFP_BERT| physics list and turned on the simulation of  short-lived particles.

\subsection{Algorithm}
Conceptually, the biasing scheme turns out to be straight-forward:
\begin{enumerate}[label=S.\arabic*,ref=S.\arabic*]
    \item The event begins by shooting the primary proton, which at some point interacts inelastically and produces secondary particles of interest (mesons or photons)
    \item \label{Algo:StepCloning} When the first interesting particle reaches the end of its first step, we add an identical particle to the stack of secondaries. Here the original is marked as "analogue" (a) and the clone is marked as "biased" (b). The kinematics and starting point of (b) must be the same as those of (a).
    \item The simulation of the original particle continues until it is destroyed or it leaves the world volume. Any time (a) creates other interesting particles, \ref{Algo:StepCloning} is applied to them.
    \item \label{Algo:StepXS} The simulation of the biased particle starts. At each step the cross-sections of processes that would kill (b) without producing muons are set to 0. At the same time, the interaction lengths of processes leading to muons in the final state are set to
    \begin{equation}
        \label{eq:Cross-section-biased}
        \lambda_{b}= \lambda_{a}\left[1-\exp(-l/\lambda_{a})\right]
    \end{equation}
    where $\lambda_a$ is the analogue interaction length of the process and $l$ is the distance between the current position of the particle and the projection along its current momentum on the plane at which the muons must be scored. Only this step modifies the weights (probabilities) carried by the biased particles.  These weights are automatically computed, stored and propagated to daughters by \verb|GEANT4|.
    \item \label{Algo:FinalStep} Whenever (b) produces another interesting particle, the secondary particle is marked as analogue, but further cloning is prevented.
\end{enumerate}
When the simulation is done, there will be a mixture of biased and analogue muons in the output, but all of the biased ones will have weights strictly lower than 1. This criterion can be used to discard the analogue component, which would otherwise spoil the statistical power of the biased sample. 

Having set the informal description of the algorithm, we now turn to the concrete implementation in \verb|GEANT4|. We refer to the classes of the GEANT4 biasing framework~\cite{Geant4Manual}.  

We see that flagging the original
particles and their clones is essential for the scheme to work. One can achieve this by deriving from the pure virtual class \verb|G4VUserTrackInformation| to store the flag and possibly other pieces information (e.g. the PDG encoding of the mother particle of each track). 

The cloning of each track is a "non-physics" biasing operation for which a \verb|G4VBiasingOperation| class is needed such that \verb|GEANT4| can handle everything swiftly. For our purposes, this class needs to implement the pure virtual methods \verb|DistanceToApplyOperation| (used to decide whether or not a given particle must be cloned) and \verb|GenerateBiasingFinalState| (used for the actual cloning operation). The former should always return \verb|DBL_MAX|, but set the \verb|G4ForceCondition| to \verb|true| only for the first step of each interesting track. For the remaining steps, the condition is set to \verb|false|. In the \verb|GenerateBiasingFinalState| we create a new \verb|G4Track| by copying the one of the incoming particle. We then assign as \textit{current} momentum and position of this clone, the momentum and position of the original track \textit{at its creation}, which can be significantly different from the ones \textit{at the end of the current step}. An object of type \verb|G4VParticleChange| is then created, using only the clone.

The next step is to ensure that all the cross-sections are handled as in point (S.4) above. \verb|GEANT4| has a built-in operation, \verb|G4BOptnChangeCrossSection|, which allows the user to modify any cross-section by any factor at each step during the simulation of a particle. 

The biasing interface of \verb|GEANT4| additionally requires a \verb|G4VBiasingOperator| class to handle all these operations. \verb|ProposeOccurenceBiasingOperation| acts only on the clones by making null all cross-sections of processes not leading to muons. The other mandatory method \verb|ProposeNonPhysicsBiasingOperation|, on the other hand, acts only on original tracks through the cloning operation defined above. 

\begin{figure*}
    \centering
\begin{tikzpicture}[
roundnode/.style={circle, draw=green!60, fill=green!5, very thick, minimum size=7mm},
squarednode/.style={rectangle, draw=red!60, fill=red!5, very thick, minimum size=5mm},
>=stealth,
node distance=2.cm,
on grid,
auto
]

\node[rectangle, draw=blue!60, fill=blue!5, very thick,text width=1.5cm,text centered]
    (piPlus) { $\pi^{+} (a)$};

\node[rectangle, draw=red!60, fill=red!5, very thick, text width = 3.0cm, text centered, below =  2.5cm of piPlus] (splitOp2) {Cloning operation as above.};

\node[rectangle,draw=black!60, very thick,text width=1.5cm,text centered, right=of piPlus]
    (unInteresting) {X...};

\node[rectangle,draw=black!60, very thick,text width=1.5cm,text centered, below right= 3.0cm and 1.0 cm of unInteresting]
    (piZero) {$\pi^{0}$ $(b)$};

\node[rectangle,draw=black!60, very thick,text width=1.5cm,text centered, below left= 2.0cm and 1.2 cm of piZero] (gamma1) {$\gamma$ $(b)$};

\node[rectangle, draw=black!60, very thick,text width=1.5cm,text centered, below right=2.0cm and 1.2 cm of piZero] (gamma2) {$\gamma$ $(b)$};

\node[rectangle,draw=black!60, very thick,text width=1.5cm,text centered, below right=0.5cm and 3cm of piZero] (neutrino) {$\nu$ $(b)$};

\node[rectangle, draw=green!60, fill=green!5, very thick,text width=1.5cm,text centered, above right= 0.5cm and 3cm of neutrino]
    (muon) {$\mu^{+}$ $(b)$};


\node[rectangle, draw=red!60, fill=red!5, very thick, text width=4cm, text centered, above left =2.5 cm and 1 cm of unInteresting] (inelastic) {Inelastic\\ \small{internally by GEANT4}};

\node[rectangle, draw=red!60, fill=red!5, very thick, text width=4cm, text centered, below right  =1.0 cm and 4.5 cm  of unInteresting] (decay) {Decay\\ \small{internally by GEANT4}};

\node[rectangle, draw=black!60, very thick,text width=1.5cm,text centered, above = of inelastic] (analogueAfterClone) { $K^{+} (a)$};

\node[rectangle,dashed, draw=black!60, very thick,text width=1.5cm,text centered, above = 5.4 cm of decay] (cloneAtCreation) { $K^{+} (b)$};
  
\node[rectangle, draw=red!60, fill=red!5, very thick, text width = 7.0cm, text centered, above right = 2cm and 2.5cm of analogueAfterClone] (splitOp) {Cloning operation. \\ ProposeNonPhysicsBiasingOperation \\ $\to$ GenerateBiasingFinalState};

\node[rectangle, draw=blue!60, fill = blue!5, very thick,text width=2.5cm,text centered, above = 5 cm of splitOp] (analogueInit) { $K^{+} (a)$};

\path [line] (analogueInit) -- node [text width = 2cm, midway,right, text centered] {\small{First step $w=1$}} node [text width = 7.0cm, midway, below, sloped, text centered] {\small{DistanceToApplyOperation\\ G4FoceCondition $=$ forced;\\ return DBL\_MAX}}(splitOp);
\path [line] (piPlus) -- (splitOp2);

\path [line] (splitOp.south)-- node[text width = 1.5cm, midway, above,sloped, text centered] {\small{$w=1$}} (analogueAfterClone.north);
\path [line] (splitOp.south)-- node[text width = 1.5cm, midway, above,sloped, text centered] {\small{$w=1$}} (cloneAtCreation.north);

\path [line] (analogueAfterClone.south) -- node[text width = 2cm, midway, left, text centered] {\small{Analogue physics $w=1$}} (inelastic.north);
\path [line] (cloneAtCreation.south) -- node[text width = 2cm, midway, right, text centered] {\small{Biased physics; $N$ steps $w_{i+1}<w_{i}$}} node[text width = 5.0cm, midway, below, sloped, text centered] {\small{ProposeOccurenceBiasing \\
$\to$ G4BOptnChangeCrossSection\\
cross-sections as in Eq.~\ref{eq:Cross-section-biased}}} (decay.north);

\path [line] (inelastic.south) -- node[text width = 1.5cm, midway, above, text centered, sloped] {\small{$w=1$}} (piPlus.north);
\path [line] (inelastic.south) -- node[text width = 1.5cm, midway, above, text centered, sloped] {\small{$w=1$}} (unInteresting.north);

\path [line] (decay.south) -- node[text width = 1.5cm, midway, above, text centered, sloped] {\small{$w=w_{N}$}} (piZero.north);
\path [line] (decay.south) -- node[text width = 1.5cm, midway, above, text centered,sloped] {\small{$w=w_{N}$}} (neutrino.north);
\path [line] (decay.south) -- node[text width = 1.5cm, midway, above, text centered,sloped] {\small{$w=w_{N}$}} (muon.north);

\path[line] (piZero.south)-- node[text width = 1.5cm, midway, above, text centered,sloped] {\small{$w=w_N$}}(gamma1.north);
\path[line] (piZero.south)-- node[text width = 1.5cm, midway, above , text centered, sloped] {\small{$w=w_N$}}(gamma2.north);

\end{tikzpicture}
\caption{Example of the biasing algorithm showing operations applied on analogue ($a$) and biased ($b$) particles. Particles to be cloned at the end of their first step are represented by blue rectangles. Solid black rectangles indicate particles undergoing analogue physics processes, but no cloning. Dashed rectangles are used for particles with modified cross-sections along their path. Particles to be kept at the end of the simulation are shown in green rectangles. The evolution of the weights ($w$) is also shown.}
\label{fig:flowchart}
\end{figure*}
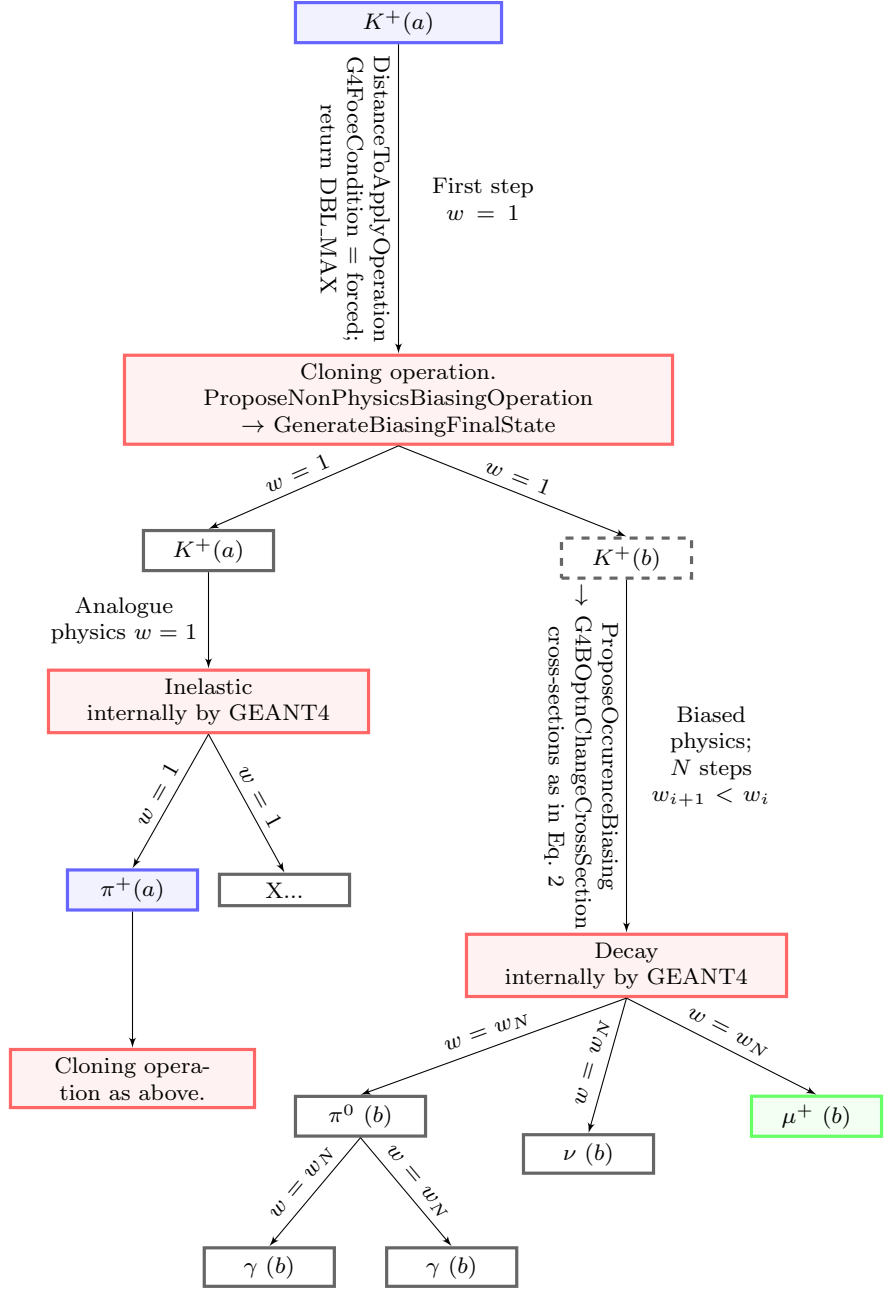

Fig.~\ref{fig:flowchart} illustrates the steps above applied to a $K^{+}$.

Before showing the comparison of this method with the analogue simulation, we note a few aspects that users of this method should be aware of. 

Step \ref{Algo:FinalStep} of the algorithm essentially removes muons of very small weights, that would be in any case discarded in a real application. Eq.~\ref{eq:Cross-section-biased} is technically incorrect for charged particles as it assumes the analogue interaction length to be constant along the step. However, for above-GeV energies this limitation becomes irrelevant, as we will show in the next section. 

One of the caveats of this biasing scheme is that its results are reliable only in the approximation that at most 1 muon/event reaches the scoring plane in the analogue setup. While this is not necessarily the case, we will show that for the simplistic geometry in our simulation, the analogue and biased samples agree very well.

Finally, it is worth drawing some attention to the resulting weights. The weight distribution has an average of $\simeq3\times10^{-4}$ and an RMS of $\simeq2\times10^{-3}$. The transformations in \ref{Algo:StepXS} produce weights of very small values (less than $10^{-20}$), in particular for muons generated by photon conversion. At the same time, about $1\%$ of the  weights are above 0.01, with a tail extending up to 1. When using the MC sample for background generation, extremely low weights induce an efficiency loss, while the muons with very large weight might affect the fluctuations. Possible solutions to these issues depend on the wanted application. One option would be to define a weight window, so that muons above a maximum threshold are split and muons below a minimum threshold undergo a Russian Roulette (see~\cite{Geant4Manual} for details). The choice of the weight window, the split multiplicity, and the Russian Roulette survival weight would be highly dependent on the application and would have to be optimised by the users.

The weight issues can be mitigated by choosing a constant  enhancement factor for the processes leading to muons in the final state. Instead of setting the cross-sections to zero for other processes, one simply kills the clones and their daughters if the end process does not generate muons. The enhancement factor should depend on the mother particle species, not to alter the muon composition. With such an approach, the low weights are completely eliminated. Still one might impose some weight window if large fluctuations are seen to be induced by the few events in the high-weight tail. Again, all the involved parameters would have to be chosen by users and tuned according to the application. We stress that there is a natural trade-off between simulation efficiency and statistical power when one deals with biased MC samples. The algorithm proposed in this paper serves the primary purpose to allow the exploration of the full phase space of muons coming from proton interactions in thick absorbers. Reducing statistical fluctuations is left to the users.
\section{Results}

Using the setup described in the previous section we have obtained $3.7\times 10^{-3}~\mu^+$/POT in the analogue sample and $15~\mu^+$/POT in the biased sample, in the entire momentum spectrum. As opposed to the increase in statistics by more than three orders of magnitude, the CPU time overhead per event introduced by the biasing is roughly equivalent to the CPU time needed for an analogue event. On a \verb|CentOS 7| machine with Intel Xeon Gold 6230R, simulating $10^4$ events requires about 350~s in the analogue setup and 720~s in the biased one. The hadronic shower is simulated faster in volumes which are somewhat homogeneous (such as the absorber in our simulation) and this implies that adding supplementary particles, though not undergoing heavy processes, increases the processing time significantly. The net increase in statistics one can achieve using our biasing scheme is a factor around 2000, while for more complex geometries, this gain is expected to increase somewhat.

\begin{figure}[h]
    \centering
    \includegraphics[width=\columnwidth]{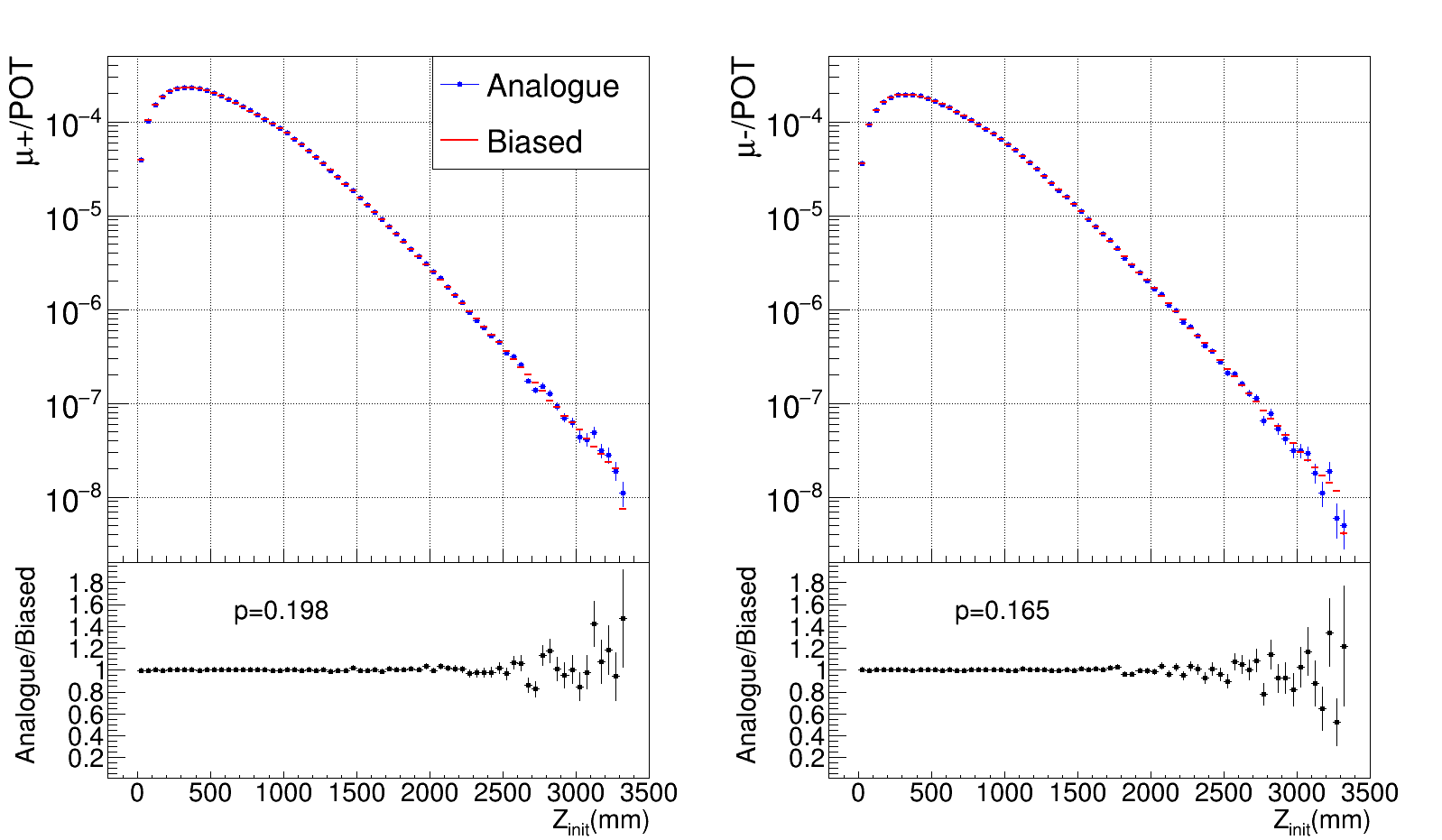}
    
    \includegraphics[width=\columnwidth]{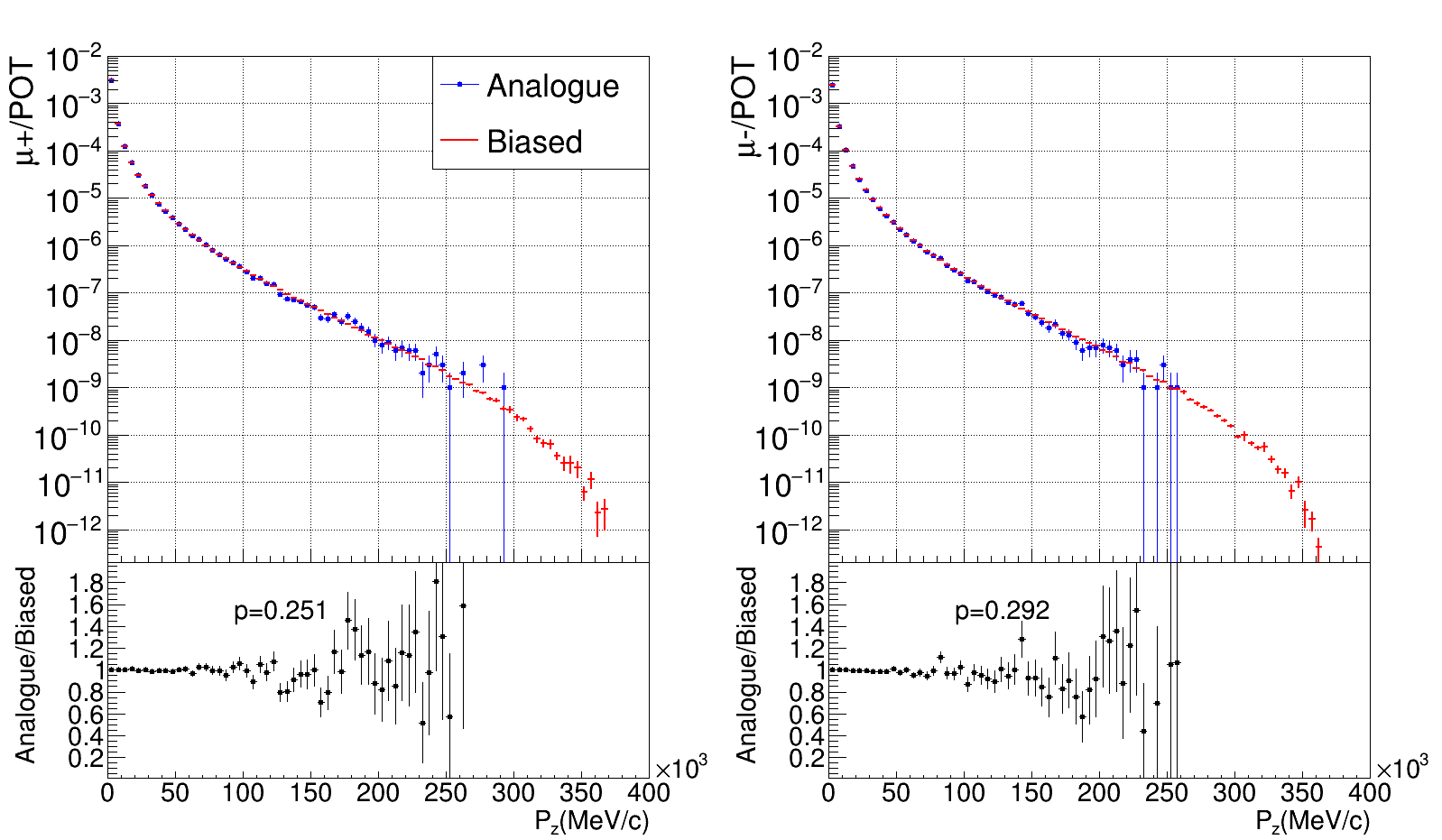}
    \caption{$z$-coordinate of muon emission (top) and $z$-component of muon momentum at the scoring plane (bottom) in the analogue (symbols, blue) and biased (continuous, red) samples with positive particles plotted on the left column and negative ones on the right column. Ratio plots, together with the $p$-value from the $\chi^{2}$ test for the analogue and biased distributions, are also shown.}
    \label{fig:PzInitZ}
\end{figure}

\begin{figure}[h]
    \centering
    \includegraphics[width=\columnwidth]{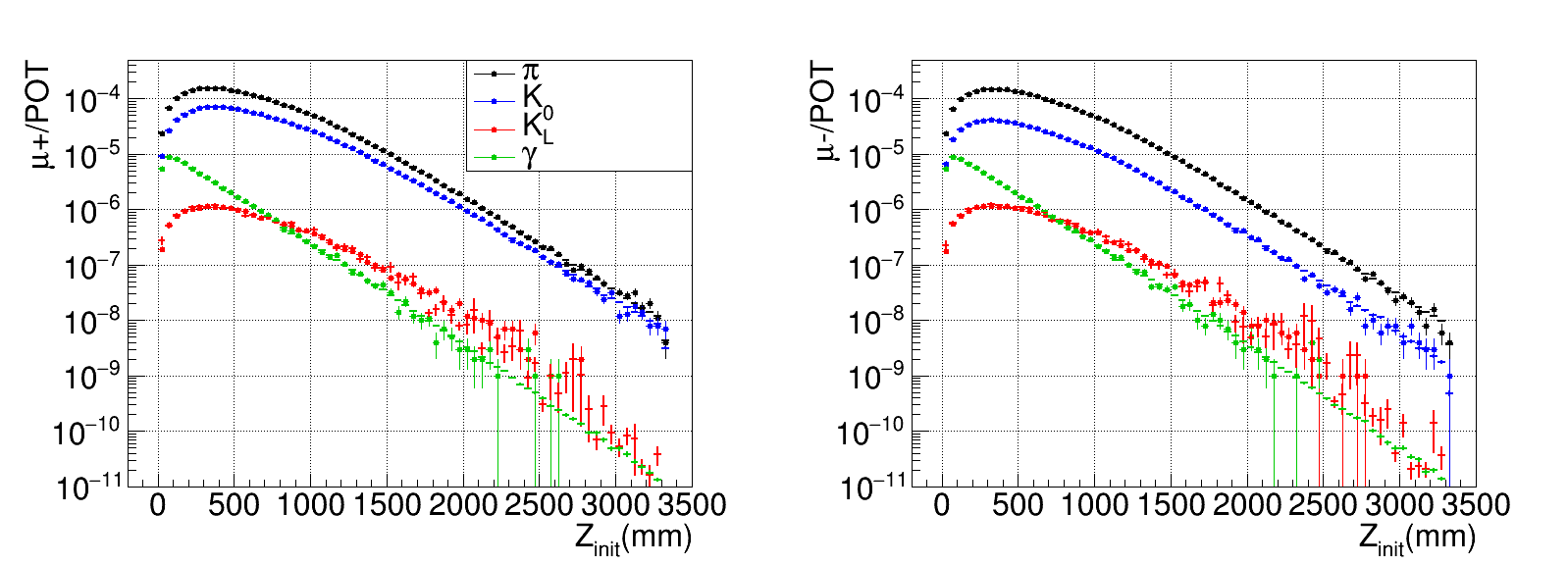}
    
    \includegraphics[width=\columnwidth]{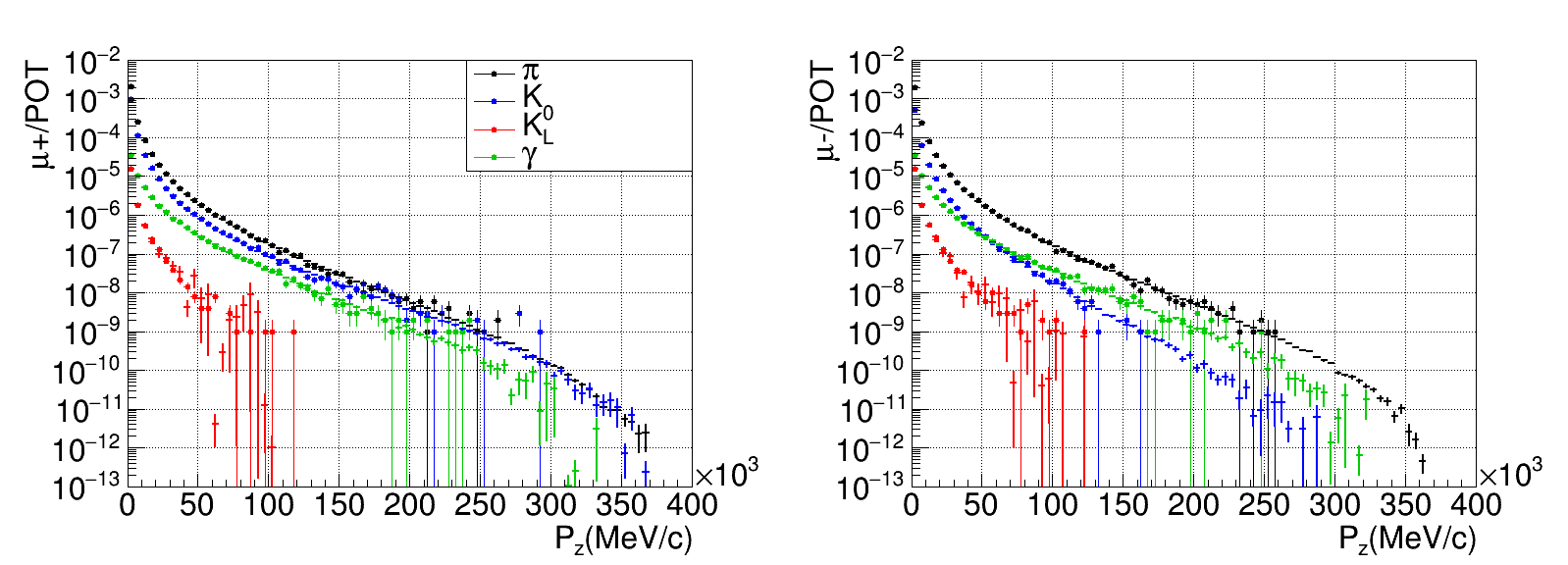}
    \caption{Same as in Fig.~\ref{fig:PzInitZ}, but with contributions split by mother particle.}
    \label{fig:PzInitZvsMother}
\end{figure}

Fig.~\ref{fig:PzInitZ} shows an excellent agreement between analogue and biased simulations in both the $z$-component of the muon momentum and the $z$-coordinate of the muon origin. The analogue sample starts to be poorly populated for all practical purposes at $P_{z}\simeq 150~\mathrm{GeV/c}$, while this not the case for the biased sample even at momenta around $300~\mathrm{GeV/c}$. The ratio plots show remarkable stability over the domain in which the analogue sample is adequately populated. We have also performed the $\chi^2$ test between the analogue and biased distributions. The ranges considered for the test are $Z_{\mathrm{init}}<3000\mathrm{mm}$ and $P_{z} < 200 \mathrm{GeV/c}$ respectively. The resulting $p$-values assure that the samples are equivalent to a very large extent. In Fig.~\ref{fig:PzInitZvsMother}, the same distributions are split by mother particle. One can see that the agreement still holds nicely, even for sub-leading sources ( $K_{L}^{0}$s and $\gamma$s).

In order to ensure the compatibility between the two samples, besides the momentum and vertex we must also check the agreement in terms of position and direction of the muons scored at the downstream face of the absorber. To this purpose, we stored the 2D distributions of $P_{x}/P_{z}\equiv x^\prime$ vs $x$ and, similarly, for $y^\prime$ vs $y$ in $P_{z}$ bins of $5~\mathrm{GeV/c}$ width. From these histograms, we then extracted the standard deviations $\sigma_{x,x^\prime,y,y^\prime}$ together with their errors. 
\begin{figure}[h]
    \centering
    \includegraphics[width=\columnwidth]{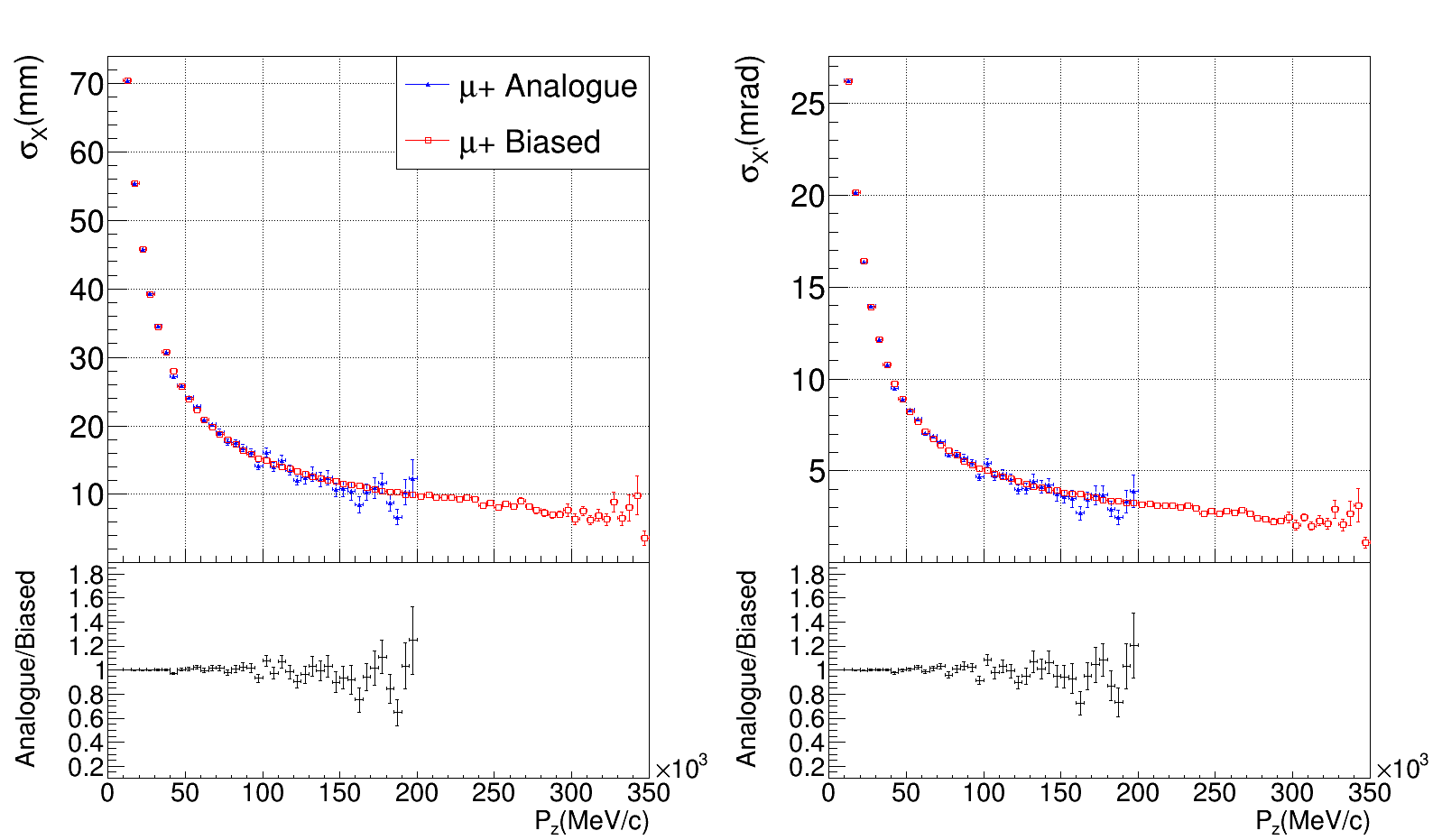}
    
    \includegraphics[width=\columnwidth]{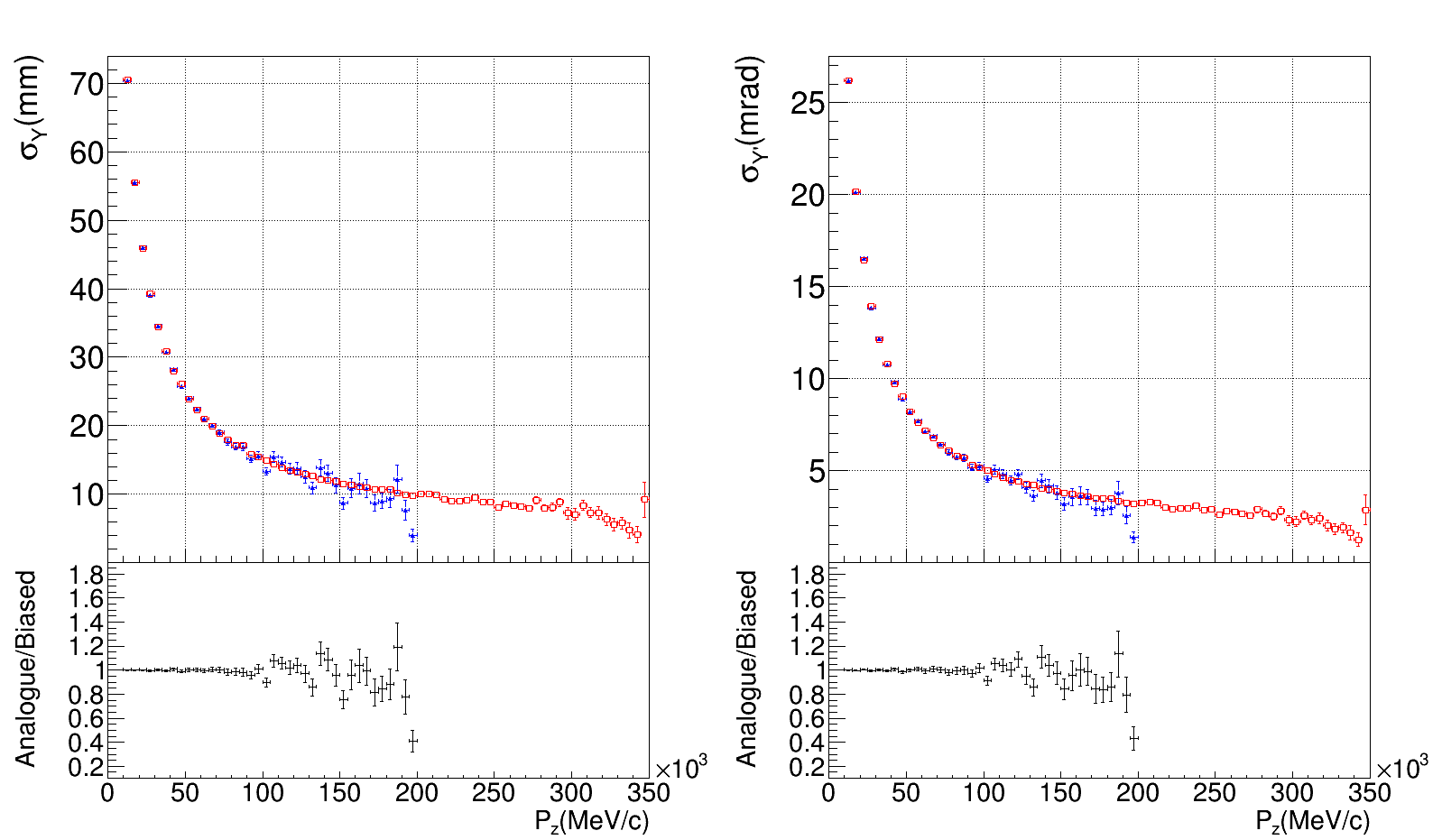}
    \caption{Standard deviation of the $\mu^+$ position $x,y$ (left panels) and muon direction angle $x^\prime,y^\prime$ (right panels) at the downstream face of the absorber as a function of the $\mu^+$ momentum. Distributions for $x$ and $x^\prime$ ($y$ and $y^\prime$) are shown in the top (bottom) panels. The analogue (biased) distribution is shown by the  blue filled (red open) dots. The ratios between the analogue and biased distributions are also shown.}
    \label{fig:muonPlus_kinematics}
\end{figure}
\begin{figure}[h]
    \centering
    \includegraphics[width=\columnwidth]{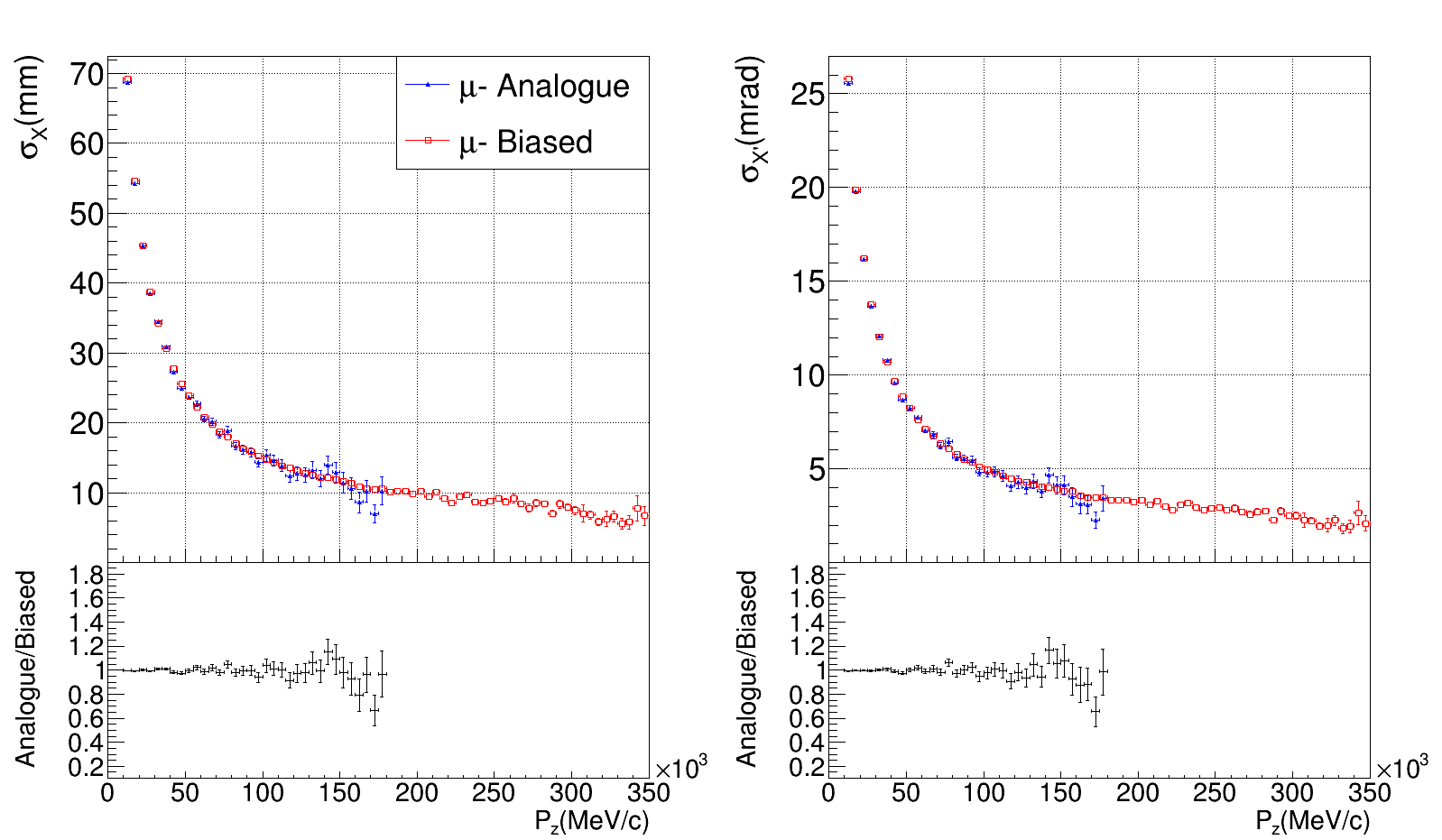}
    
    \includegraphics[width=\columnwidth]{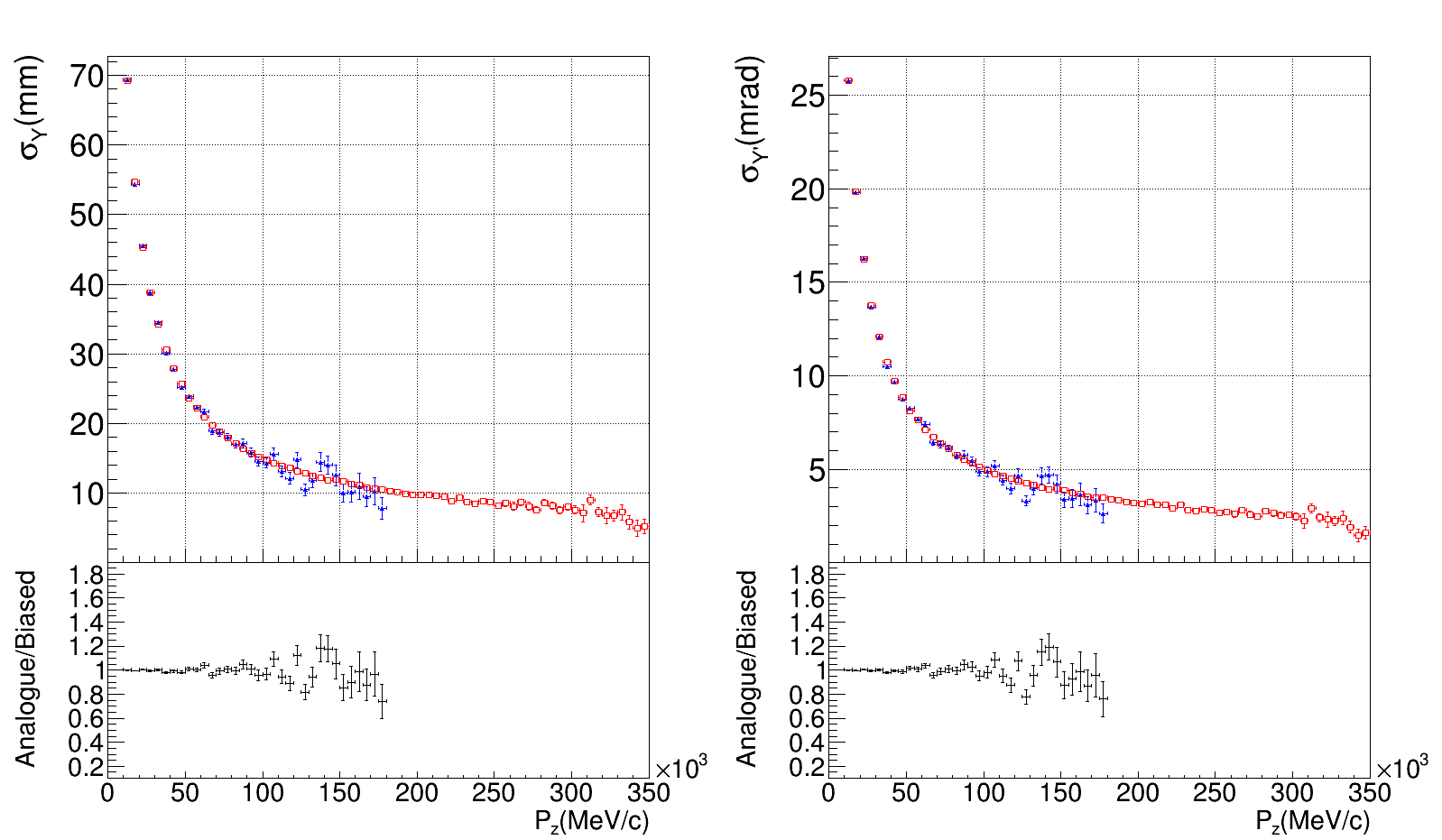}
    \caption{Same as in Fig.~\ref{fig:muonPlus_kinematics}, but for negative muons}
    \label{fig:muonMinus_kinematics}
\end{figure}
Figs.~\ref{fig:muonPlus_kinematics} and \ref{fig:muonMinus_kinematics} show these quantities as function of $P_{z}$ for positive and negative muons. The analogue distributions end at $\simeq 200~\mathrm{GeV/c}$ due to insufficient statistics (less than 100 entries in each $P_{z}$ bin). These last two plots ensure that the analogue and biased samples are compatible.\\

In summary, in this paper we have outlined a method to achieve an $\mathcal{O}(10^3)$ fold improvement in the statistical power for the production of muon halos in beam-dump experiments.
The method has been validated using a simplified mock-up version of the NA62 beam dump. We expect that this method can be adapted in a straightforward fashion to a detailed description of the NA62 experiment, as well as to experiments with similar setup. The method can be directly applied to the simulation of neutrino beams, as well. 
Our results constitute a significant step forward in the aim to determine the expected background to searches for new-physics feebly-interacting particles from first principles. 

\section*{Acknowledgements}
  SG acknowledges the support of the Institute of Atomic Physics through the CERN-RO Project no. 10/10.03.2020.
  BD and EM acknowledge support through the European Research Council under grant ERC-2018-StG-802836 (AxScale project).
  We also would like thank Simone Schuchmann for useful discussions on the presented work.

\end{document}